\documentclass[11pt]{article}
\usepackage{geometry}                
\usepackage{hyperref}
\usepackage{subfigure}
\usepackage{graphicx}
\usepackage{amssymb}
\usepackage{epstopdf}
\usepackage{amsmath}
\title{A Mathematical Framework for Agent Based Models of Complex Biological Networks
\thanks{ 
This work was supported by a grant from the U.S. Army Research Office. 
The authors are grateful to the National Institute for
Mathematical and Biological Synthesis, which is sponsored by the National Science
Foundation, the U.S. Department of Homeland Security, and the U.S. Department
of Agriculture through NSF Award \#EF-0832858, with additional support from The
University of Tennessee, Knoxville. We have benefited greatly from the workshop
``Investigative Workshop on Optimal Control and
Optimization for Individual-based and Agent-based Models" held there in December 2009.
The authors are grateful to all the participants of this workshop for stimulating 
discussions and insights. In particular, the authors thank Volker Grimm,
Virginia Pasour, and Grigoriy Blekherman for
helpful comments on an earlier draft of the manuscript.}
}
\author{Franziska Hinkelmann$^{a,b}$, David Murrugarra$^{a,b}$, Abdul Salam
Jarrah$^{b,c}$, \\Reinhard Laubenbacher$^{a,b,*}$}

\begin{document}
\maketitle

{\footnotesize
     \centerline{$^a$Department of Mathematics,
      Virginia Polytechnic Institute and State University,}
  \centerline{Blacksburg, VA 24061-0123, USA}
}
{\footnotesize
    \centerline{$^b$Virginia Bioinformatics Institute,
      Virginia Polytechnic Institute and State University,}
  \centerline{Blacksburg, VA 24061-0477, USA}
}
{\footnotesize
  \centerline{$^c$Department of Mathematics and Statistics,
  American University of Sharjah,}
  \centerline{Sharjah, United Arab Emirates}
}

{\footnotesize
   \centerline{$^*$Corresponding author: reinhard@vbi.vt.edu}
}

\begin{abstract}
Agent-based modeling and simulation is a useful method to study biological phenomena 
in a wide range of fields, from molecular biology to ecology. Since there is
currently no agreed-upon standard way to specify such models it is
not always easy to use published models. Also, since model
descriptions are not usually given in mathematical terms, it is difficult to 
bring mathematical analysis tools to bear, so that models are typically
studied through simulation. In order to address this issue, Grimm et al. proposed
a protocol for model specification, the so-called ODD protocol, which
provides a standard way to describe models. This paper proposes an addition
to the ODD protocol which allows the description of an agent-based model
as a dynamical system, which provides access to computational and theoretical
tools for its analysis. The mathematical framework is that of algebraic models,
that is, time-discrete dynamical systems with algebraic structure. It is shown
by way of several examples how this mathematical specification can help with
model analysis. This mathematical framework can also accommodate other
model types such as Boolean networks and the more general logical models, 
as well as Petri nets. 
\end{abstract}

\section{Introduction} 

The arsenal of modeling tools in mathematical biology has grown to include
a spectrum of methods beyond the traditional and very
successful continuous models with the introduction of Boolean network models
in the 1960s and the more general so-called logical models in the
1980s \cite{thomas}. Since then other methods have been added, 
in particular Petri nets (see, e.g., \cite{petri_nets}) as models for metabolic
and molecular regulatory networks. More recently, agent-based, or individual-based
models, long popular in social science, have been used increasingly 
in areas ranging from molecular to population biology. Discrete models
such as these have many useful features. Qualitative models of molecular
networks such as logical models, do not require kinetic parameters but
can still provide information about network dynamics and serve as tools for hypothesis generation. 
Agent-based models can capture the fact that in some biological systems, such
as a growing tumor, system dynamics emerges from interactions at the local level,
such as cell-cell interactions in the case of a tumor. Discrete models also tend to be
more intuitive than models based on differential equations, so they have added appeal
for researchers without a strong mathematical background. 

The flip side of the coin is the relative lack of mathematical analysis tools for discrete models. 
While methods like bifurcation, sensitivity, and stability analysis are available for differential 
equations models, the principal tool in the discrete case is simulation. While this is very
effective for small models, it becomes impossible for larger models, since the size
of the phase space is exponential in the number of variables in the model. Thus, problems
like the identification of steady states for a Boolean network model becomes problematic
once the model contains many more than 20 or 30 nodes, unless one makes use of high performance
computation capabilities. An added complication is the heterogeneity of the different discrete
model types so that tools developed for one type are unlikely to apply to another one. 

One possible approach to this problem is to find a mathematical framework that is general
enough so that most or all types of discrete models can be formulated within this framework
and is rich enough to provide practically useful theoretical and computational tools for model
analysis. This approach was taken in \cite{Alan:Bioinf2010}, where it was shown that any logical model
and any $k$-bounded Petri net can be translated into a time-discrete dynamical system over a 
finite state space. The transition function can be described in terms of polynomial functions. 
This makes model analysis amenable to the computational tools and theoretical results of 
computer algebra, a theoretically rich area that has taken advantage in the last decade of increasingly
powerful symbolic computation capabilities. In this setting, the computation of steady states
of a model, for instance, turns into the problem of solving a system of polynomial equations, for 
which there are several good software implementations. 
In this paper we show that one can make a similar translation for many agent-based models, thereby
covering a large part of the discrete models available in the literature. One added benefit of this 
common mathematical framework is that one can use it to easily compare models of different types. 

Many complex biological systems can be modeled effectively as multiagent systems in which
the constituent entities (agents) interact with each other. For instance, processes unfolding
in a non-homogeneous spatial environment can be modeled in this way, or processes
that are inherently discrete, such as individual immune cells interacting with each other 
in a volume of tissues. Examples include \cite{EBV3, episims, nfkb-agent, Wang:2009rm, Peer:2005}. 
Often, the models include large numbers of agents that can be in one
of finitely many different states and interact with each
other and their environment based on a set of deterministic or stochastic rules. The
global dynamics emerge from
the local interactions among the agents. The advantage of increased realism of
agent-based models (ABMs) is counter-balanced by the relative lack of mathematical tools for
their development and analysis. 

One key obstacle is
the lack of a formal description of ABMs in a way that makes them amenable
to mathematical tools for analysis and optimal control. An important step in this
direction is the work of Grimm et al., which provides a protocol for model specification. 
In \cite{Grimm2006115} the authors point out that agent or individual based
models are ``more difficult to analyze, understand, and communicate''
than traditional analytical models because they are not ``formulated in the
general language of mathematics.'' They proceed to develop the so-called ODD protocol
for the specification of such models. The basic mathematical nature of many ABMs is that of a time-discrete dynamical system on a finite state space
(either deterministic or stochastic). A state of the model can be specified as a 
vector of values, one for each of the model variables. In addition, a function, either
deterministic or stochastic, is specified that transforms a given model
state into another state. Model dynamics is generated by iteration of this function. 
There are other model types, such as discrete event simulations, that can also be
viewed in this framework. Even hybrid models that contain continuously varying
quantities, such as temperature, can sometimes be treated as discrete models in practice.
The ODD protocol provides essentially a standard template
for specifying the state space of the model and the update function. 

Little is gained in terms of mathematical power by viewing an ABMs as a function from
the set of states to itself, without any additional mathematical structure, however. 
And it would still be difficult to verify, in many cases, whether the update function has been
specified completely and unambiguously, an important aim of the ODD protocol. 
Both problems could be remedied by the
introduction of additional mathematical structure that provides access to mathematical
tools and theoretical results, and which at the same time respects the fundamental property of ABMs that global
dynamics emerges from local interactions. And the additional mathematical structure should
be ``benign,'' in the sense that it introduces few or no mathematical artifacts into the model
properties, in particular its dynamics. Furthermore, it should be computationally tractable, allowing easy model comparison, for instance. The addition
of such structure to the ODD protocol in a way that takes a model specified in the current
ODD protocol and translates it automatically into a mathematical object would place
little burden on the user, while giving access to mathematical tools. In this paper we
propose such a mathematical structure and demonstrate its use via some
examples.

A natural way to approximate ABMs by state space models that
are grounded in a richer mathematical theory and satisfies the constraints discussed above is to 
construct an algebraic model specification, that is, a discrete time, discrete state
dynamical system whose state space represents exactly the dynamic properties of the ABMs. 
Algebraic models can be described by 
polynomial functions over finite fields, which provides access to the rich algorithmic theory of computer
algebra and the theoretical foundation of algebraic geometry. 

We demonstrate the added value that is gained from such a mathematical description through a collection
of examples. The first example illustrates the fact that it is easy to check by comparing polynomials
whether two different implementations of the
same model are identical, using a published ABM of butterfly migration \cite{Peer:2005}. 
The second example consists of an 
epidemiological model in the form of a cellular automaton. Here one can use theoretical mathematical results
to determine all periodic points of the model and their period without resorting to simulation.
Finally, we show how to compute all steady states of Conway's Game of Life using computer algebra algorithms
for the solution of systems of polynomial equations. In a forthcoming paper we will illustrate the use of the
mathematical framework proposed here for the purpose of designing optimal control methods for agent-based
models.

\section{Algebraic Models}
The basic idea underlying our approach is very similar to the idea that allowed geometers
to move from synthetic geometry to analytic geometry, namely the introduction of a coordinate system. 
That is, we need to impose an algebraic structure of addition and multiplication on the set of 
possible states of the model variables, so that we obtain a field. (This assumption has
long been made in the case of Boolean networks, where the choice of underlying field is the
Galois field $\mathbb F_2 =\{0, 1\}$.) This is possible whenever the
number of states for a given variable is a power of a prime number. In practice, it is easy to 
accomplish that all variables take values in the same finite field $\mathbb F$, 
either by choosing an appropriate number of levels for a given variable at the outset
or by introducing duplicates of one or more states. As with coordinate systems on real-valued spaces,
there will generally be several different choices that all lead to equivalent outcomes. 
Once we choose such an algebraic structure $\mathbb F$, then the set function 
description of an ABM turns into a mapping between vector spaces over the finite field $\mathbb F$,
which can be described in terms of polynomial coordinate functions. We briefly describe this
class of dynamical systems and then provide a detailed description of how to translate
an ABM specified in the ODD protocol into such a polynomial dynamical system.

Let $x_1, \ldots, x_n$ be a collection of variables, which take
values in $\mathbb F$. The variables
represent the entities in the system being modeled and the
elements of $\mathbb F$ represent all possible variable states. 
Each variable $x_i$ has
associated to it a ``local update function'' $f_i:
\mathbb F^n \longrightarrow \mathbb F$, where ``local" refers to the fact that
$f_i$ takes inputs from the variables in the ``neighborhood" of
$x_i$. Here, ``neighborhood" refers to an appropriately defined directed
graph encoding the variable dependencies. These functions assemble to a
dynamical system
\[
f =(f_1,\dots,f_n): \mathbb F^n \longrightarrow \mathbb F^n,
\]
with the dynamics generated by iteration of $f$. Iteration can be performed by
updating the variables synchronously or asynchronously. 

The dynamics of $f$ is usually represented as a directed graph on
the vertex set $\mathbb F^n$, called the \emph{state space} of $f$. There
is a directed edge from $\mathbf v \in \mathbb F^n$ to $\mathbf u \in \mathbb F^n$ if
and only if $f(\mathbf v) = \mathbf u$. It is easy to show \cite{LN} that
any local function $f_i :\mathbb F^n \longrightarrow \mathbb F$ can be expressed as a polynomial
in the variables $x_1,\dots, x_n$. 
This observation has many useful consequences, since polynomial
functions have been studied extensively and many analytical tools
are available. 

\begin{figure}[ht]
\centerline{ \raise10pt\hbox{ \framebox{
\includegraphics[width=0.1\textwidth]{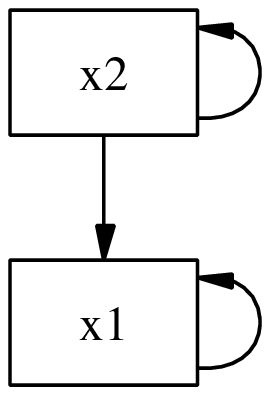}
} } \qquad \qquad \framebox{
\includegraphics[width=.4\textwidth]{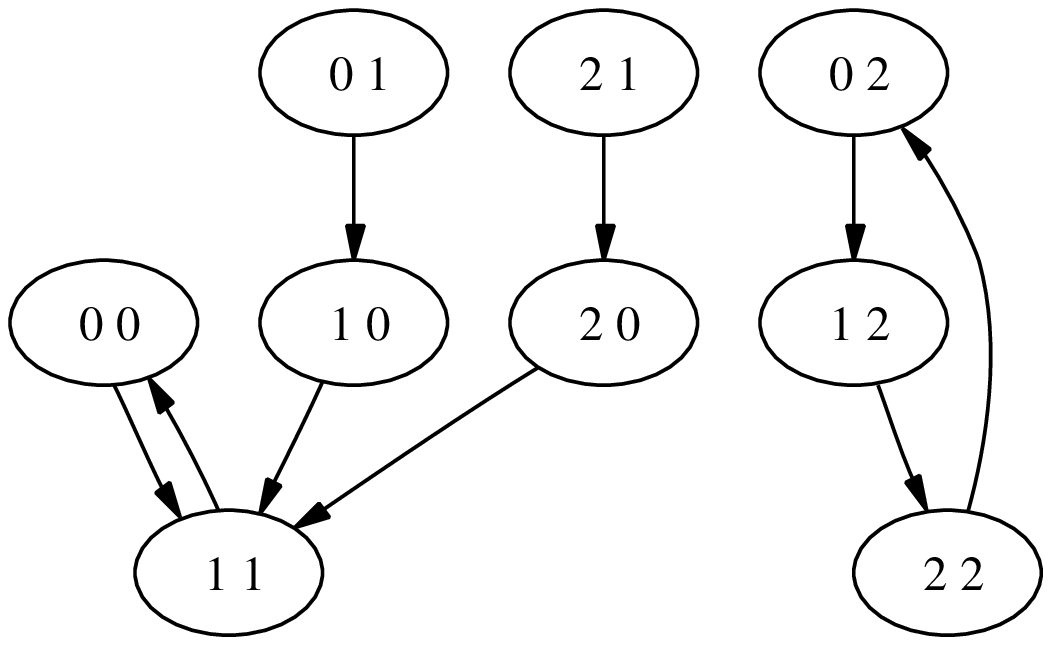}
} } \caption{The dependency graph (left) and the state space
$\mathcal{P}(f)$ (right) of the polynomial dynamical system in the above example.} 
\label{fig1}
\end{figure}

We discuss a simple example.
Let $f : \mathbb F_3^2 \longrightarrow \mathbb F_3^2$ be given by $f(x_1,x_2)
= (1-x_1x_2, 1+2x_2)$. The state space of $f$ has two components,
containing two limit cycles: one of length two and one of length
three. See Figure \ref{fig1} (right). The dependency relations
among the variables are encoded in the dependency graph in Figure
\ref{fig1} (left).

It is discussed in \cite{encyclop} that the framework of algebraic models is particularly suitable
for the study of agent-based simulations, since many agent-based simulations naturally
map to this mathematical setting. Furthermore, it grounds the investigation firmly in the
mathematical fields of dynamical systems and polynomial algebra, both of which provide
a rich set of tools and concepts. 
\section{Polynomial Form of ODD Models} 
In \cite{Grimm2006115}, the authors propose a standard protocol, named ODD after
its key components {\emph {Overview, Design concepts, and Details}}, for describing individual
based and agent-based models. The main motivation was to better enable the communication of
such models. They state: ``We conclude that what we badly need is a standard protocol for
describing IBMs which combines two elements: (1) a general structure for describing IBMs, thereby 
making a model's description independent of its specific structure, purpose and form of 
implementation [...] and (2) the language of mathematics, thereby clearly separating verbal
considerations from a mathematical description of the equations, rules, and schedules that
constitute the model.'' In this section we address the second element and first review the key
features of the ODD protocol, using the categories from \cite{Grimm2006115}. We will make
two assumptions on the models this section applies to.
\begin{itemize}
\item
All state variables in the model are updated in discrete time steps, either explicitly in the model or 
in the way state variable updates are computed.
\item
All state variables can take on only finitely many different states. (This includes state variables
that include probabilities, etc., since in practice these are represented by only finitely many choices.)
\end{itemize}
While these assumptions exclude some models, they are satisfied for many ABMs found
in the literature.
We next describe the different parts of an ODD model and how they relate to
algebraic models.

\subsection{Purpose} \label{sec:purpose}
This part contains a verbal description of the process the model is
intended to capture and the questions one hopes to answer using the model.

\subsection{State Variables and Scales}
The term ``state variable" refers to the low-level variables that characterize the low-level entities of the model, such as individuals or spatial entities. Another class of variables to be considered are aggregated variables
such as population size or average food density. These auxiliary variables contain information that is
deduced from low-level state variables. Thus, the state variables represent the fundamental components of the 
system, the parts whose interactions create the whole. Aggregate variables contain information
about the system by aggregating information about the state variables. State variables can be grouped
according to type, e.g., individuals, spatial state variables, etc.:
$$
x_1, \ldots , x_n; y_1, \ldots , y_m; \ldots ; z_1, \ldots , z_r.
$$
Each state variable $x$ can take on values from a prescribed set $X$. For
instance, an individual could be described by the state vector $({\text {age, sex, location}})$, so that
$X$ consists of a set of triples with a mixture of numerical and symbolic entries. 
A spatial location $y$ could be described by the state vector $({\text {number of cars occupying the location,
traffic flow}})$, so that its state set $Y$ contains a set of pairs with
numerical entries. An {\emph {agent}} is described by a collection of state variables.

\subsection{Process overview and scheduling} 
This part contains a verbal description of the specific processes to be captured by the model. The
scheduling aspect is very important for our purposes. Is time modeled using discrete time steps, 
or continuous time, or both? Are there different time scales involved, e.g., slow and fast, and which variables
fall into which category? What is the update order for the different state variables, synchronous
or asynchronous, with a fixed schedule or in random order? That is, the specification needs to 
give a complete description of the update schedule for all state variables. 

\subsection{Design Concepts}
For our purposes, the important aspects addressed here are:
\begin{itemize}
\item
{\emph Adaptation.} Do the state variables change the way they interact with other state
variables, either individuals or spatial state variables, as a result of changes in their
environment? What are the environmental variables they sense?
Do they have a fitness objective that drives adaptation?

\item{\emph Interaction.} 
What are the dependencies of the state variables and what are the rules 
for their update?

\item{\emph Stochasticity.} 
Do the state variables follow deterministic or stochastic rules to update their state?
\end{itemize}

\subsection{Input}
It is necessary to specify all necessary inputs defining the state of all state variables and for
computing a state update for each variable. 

\subsection{Submodels}
This part contains a detailed description of model equations and rules, as well as all model
parameters. It should also include a detailed justification for the choices made.

\bigskip
{\emph\bf Mathematical specification.} As Grimm et al. point out, the goal has to be to obtain a
model description that is complete and as mathematical as possible. We now rephrase these
features in a more mathematical way. The fundamental components of the model are as follows.
\begin{itemize}
\item
The state variables. We will denote these by $x_1, \ldots , x_n$, without taking into account
the different groupings based on domain-specific notions such as individual or spatial entity, etc. 
\item
Each state variable $x_i$ has a set of states $X_i$ that it can be in. Thus, a state of the model
is given by an element of the Cartesian product $X = X_1\times \cdots \times X_n$. Note that
for the purpose of mathematical specification it is not important that there are different types of state
variables. This information is implicit in their set of possible states.
\item
Each state variable $x_i$ is assigned a finite collection of rules to update its state.
At each step, each state variable chooses a rule, either deterministically or stochastically, which takes as input the
states of all or some other state and environmental variables, and assigns a new state to $x_i$. The choice of rule
might involve aggregated variables and/or random choices. Note that this rule needs to provide complete
information about how to determine the new state, given any admissible input state of the state variable. For
instance, a bacterium might have several metabolic modes depending on the environment it finds itself in and
the density of other bacteria present. The update rule chosen will then depend on the relevant environmental
variables, and possibly others. 
\item
We are given a complete specification of the order in which state variables are to be updated. 
That is, to compute a new state of the model, we update some variables before others, some variables
simultaneously, and for some variables we choose a random update order. Different time scales can be
implemented by updating faster variables several times before updating slower ones. 
\end{itemize}

Observe that each rule for the update of a state variable can be expressed as a 
function $f_i: X \longrightarrow X_i$.
We can now assemble these components to represent the 
model as a time discrete dynamical system
$$
f = (f_1,\ldots , f_n): X \longrightarrow X,
$$
with dynamics generated by iteration. We describe the most general case of models that allow
state variables to evolve and choose different update rules depending on environmental conditions.
In this case, each state variable $x_i$ has associated to it a probability space $P_i$ of rules/update functions
$f_i: X \longrightarrow X_i$, which represent its different ``modes of action,"
depending on the environment. The probability distribution on $P_i$ can be computed with information
provided as part of the ODD. For instance, a state variable may choose an update function based on
the state of one or more aggregated variables that describe its environment,
such as {\it food density}, or 
the states of other state variables in its environment. For a given model update, each state variable chooses
one update function from this probability space. The details of the construction can be found in Appendix 1.
The key step in the construction is the replacement of each $X_i$ with a finite field $\mathbb F$, so that 
$X = \mathbb F^n$.

The end result is that we can now describe the ABM as a dynamical system
$$
f = (f_1,\ldots , f_n):\mathbb F^n\longrightarrow \mathbb F^n,
$$
with all $f_i\in\mathbb F[x_1,\ldots , x_n]$ polynomials. So what have we accomplished? Translating
a model specified in the ODD protocol into a polynomial dynamical system has several advantages:
\begin{itemize}
\item
Models are stored in a unified mathematical way.
\item
Ambiguities in the verbal description and incomplete information can be detected in the translation
to an equation-based model. 

\item
It is easy to implement an existing model and modify it.
\item
There exists a body of mathematical tools to analyze models, such as computing all steady states,
which amounts to solving a certain system of polynomial equations. Also, there is a framework
for optimal control in this context, which we describe in a future paper.
\item
It is easy to compare models.
\item It is easy to incorporate variables describing the global environment, such as temperature, market price, pH value, 
as external parameters into the polynomial functions. 
\end{itemize}

It is also worth mentioning that the mathematical framework we have created for ABMs in this way is
``minimal," in the sense that all we have done is to impose a mathematical structure on the state space
of the model. We have not changed or approximated the rules used to update state variables, and
we have not changed the way in which the states of the model are updated. That is, we still have
an exact representation of the ABM, but in precise mathematical terms.

Polynomials are neither intuitive nor are they simple functions. But they
provide an exact representation of the dynamics of the model that is more compact than
the state space, which is not feasible to describe for most realistic models. Any
computer algebra system can be used to analyze a polynomial system,
independent of a particular software or implementation. The polynomials can be
generated in an almost automatic way: we provide a simple script to generate
the polynomials that interpolates a given truth table \cite{Parser}, and tables are easily generated from the description of the model. We illustrate this process with an example of a model
specified in the ODD protocol, taken from the
text book \cite{GrimmRailsback2010}. 
\section{Examples}
We now show three examples, one of which demonstrates how to translate a model
specified in the ODD protocol to its algebraic representation, and the other two 
show how the algebraic representation can be used to analyze the global
dynamics of the system without simulating it. We want to point out that these
examples are meant as "proof of concept" illustrative demonstrations. A deeper
analysis of each of them to show what the algebraic language can and cannot do
is beyond the scope of this paper, and for this purpose these examples might
not be the best choices. The first model was chosen because it is a key
example in the expository book \cite{GrimmRailsback2010}. The second example
affords an easy way to demonstrate how one might use theoretical results about
algebraic models for the purpose of analyzing dynamics of models that are much
too large to study thoroughly through simulation. And the third example is a widely
known cellular automaton that has rarely been studied from the point of view of a 
dynamical system. 

\subsection{From ODD to algebraic model: Virtual Corridors of Butterflies}
We demonstrate how to formulate the algebraic description for a model given in the
ODD protocol and show how this process provides guidelines to the modeler for
including all relevant details and formulating the model such that it is suitable for
the purpose it was built for. The example we use is a model analyzing the 
emergence of virtual corridors in the movements of butterflies navigating a landscape based on
an elevation gradient \cite{Peer:2005}. This model was used in \cite{GrimmRailsback2010} as
an example of how to specify an ABM in the ODD protocol. 

We model the ``hilltopping'' behavior of butterflies, as they try to reach the
highest point in their spatial environment for mating. The model is initialized with 500 butterflies
on a landscape discretized into $150 \times 150$ square
patches. A butterfly moves with probability $q$ to the highest patch
of its 8 surrounding patches, and it moves randomly with probability $1-q$.
Initially, all butterflies start out on the same patch, and simulations are run for $1000$ iterations.

The {\it purpose}, see Section \ref{sec:purpose}, of the model is to understand what conditions lead to virtual
corridors and how the uncertainty of butterflies to sense the elevation gradient correctly
affects these virtual corridors. This clearly stated purpose forces us to use patches
as agents, i.e., use a variable $x_i$ for each patch. Butterflies are
the ``acting agents,'' so they have to be represented as
variables, too. Since the butterflies are assumed to be homogeneous
the state of a butterfly only consists of its position. We enumerate the
patches so every state corresponds to a different position. Patches differ by their elevation (which
is fixed during a simulation) and the number of butterflies on them. At the
end of a simulation, one can compare the number of butterflies on the patches
to detect virtual corridors.

\subsubsection{Algebraic Model} 
Each butterfly is a state variable, $x_1, \ldots, x_{500}$, and so is each
patch, $x_{501}, \ldots, x_{23000}$, we let $x$ denote a state of the system,
$x = (x_1, \ldots, x_{23000})$. The state of a butterfly is its
position ($1, \ldots 22500$), the state of a patch is the number of
butterflies on it ($0, \ldots, 500$). We enumerate the patches and assume that
an elevation map of the modeled area is
given. Updates are synchronous probabilistic updates with fixed probabilities. 
From the elevation map we can create 9 different tables:
the first table assigns to every patch its most elevated neighbor, the
remaining 8 tables assign to every patch its north (north-east, east,
south-east, ..., north-west) neighbor.

Since there are 22500 different states for a butterfly and we want to work
over an algebraic field, we choose $p=22501$, the next highest prime number,
as described in Appendix 1. The algebraic model is then 
a system of equations over $\mathbb F_{22501}$.

The first table has two rows. The two entries in a column, labeled $a_i$ in the
first row and $b_i$ in the second row, for $i \in \{1,\ldots, 22500\}$ indicates
that the patch adjacent to $a_i$ with the highest elevation is $b_i$.
We generate the polynomial $g_{j,1}$ that updates a butterfly
$x_j$ for $j \in \{1, \ldots, 500\}$ in the following way. 
$$g_{j,1} ( x ) = \sum _ {i = 1} ^ {22500} (1-(a_i-x_j)^{p-1}) b_i.$$
Note that for all butterflies the polynomials $g_{j,1}$ are the same because all 
butterflies have exactly the same hilltopping strategy and ability. This function
is generated as follows. The state of a butterfly is equal to the
number of the patch the butterfly is on. Therefore
$(a_i-x_j)^{p-1}$ is equal to $1$, except when $x_j = a_i$, i.e., the butterfly is
on patch $a_i$,
then it is 0. So $(1-(a_i-x_j)^{p-1})$ is equal to $0$, unless $x_j = a_i$, in which case it
is $1$, and we multiply by $b_i$. So $g_{j,1}(x)$ interpolates the first table, 
it only depends on the state of butterfly $x_j$, not on the other
butterflies or their distribution over the patches.
It is straightforward to generate $g_{j,2}(x), \ldots, g_{j,9}(x)$ for the remaining 8 tables. If
a butterfly detects the correct elevation with probability $q$, then the
probabilistic update function for butterfly $x_j$ is
$$f_{j}(x) = \left \{ 
    \begin{array}{cl}
      g_{j,1} (x) & \text{ with probability } q\\ 
      g_{j,2} (x) & \text{ with probability } \frac {1-q} 8\\ 
      g_{j,3} (x) & \text{ with probability } \frac {1-q} 8\\ 
      g_{j,4} (x) & \text{ with probability } \frac {1-q} 8\\ 
      g_{j,5} (x) & \text{ with probability } \frac {1-q} 8\\ 
      g_{j,6} (x) & \text{ with probability } \frac {1-q} 8\\ 
      g_{j,7} (x) & \text{ with probability } \frac {1-q} 8\\ 
      g_{j,8} (x) & \text{ with probability } \frac {1-q} 8\\ 
      g_{j,9} (x) & \text{ with probability } \frac {1-q} 8,
    \end{array}
  \right .$$\\
where $g_{j,i}$ interpolates table $i$. 

The functions for the state of the patches, $f_{j}$, $j \in \{501, \ldots,
23000\}$, are built the following way:
$$f_{j}(x) = \sum _{i = 1} ^{500} (1 - (x_i - j)^{p-1}).$$
As before, each summand is 1 or 0, depending on whether $x_i$ is equal to
$j$ or not, respectively. An increment of 1 is added to the state of patch $j$ whenever a
butterfly $x_i$ is in state $j$, i.e., on patch $j$. 
The algebraic system that describes the complete butterfly
model is 
$$f : \mathbb F _ {22501} ^ {500 + 22500} \rightarrow \mathbb F _ {22501} ^ {500 + 22500}, $$
$$(x_1, \ldots, x_{23000}) \mapsto (f_{1}(x), \ldots,f_{23000}(x)).$$

This example demonstrates how to formalize a model given in the ODD protocol as an
algebraic model, which fully captures all details of the butterfly hilltopping
behavior. The algebraic framework also provides further advantages to the
modeler: algebraic equations eliminate any ambiguity inherent in verbal
descriptions. The functions are specified for all possible cases, some of which
are often left out when specifying a function verbally. 

One benefit of this description is the ease of comparison of different model implementations. 
We illustrate this with an example.
In the model description in \cite{GrimmRailsback2010}, a butterfly ``moves
uphill'', i.e., ``to the neighbor patch that has the highest elevation,'' with
probability $q$. But what if a butterfly is already on the highest patch? All
neighboring patches have a lower elevation, so any movement is a
\textit{downhill}
movement. Two different interpretations are possible:
\begin{itemize}
  \item[1.] If a butterfly is on the highest patch, it stays on the same patch
    with probability $q$, to not contradict the ``move
    uphill'' instruction;
  \item[2.] butterflies can always move, even if this means that ``move uphill''
    is actually a downhill movement.
\end{itemize}
Clearly, these two interpretations lead to different update functions (which may
or may not affect the qualitative dynamics of the model). The
difference in a verbal description or implementation might be hard to notice,
but it is
easy to see the difference by comparing the resulting polynomials. Using
standard computer algebra tools, the terms of each polynomial are sorted in 
a unique way, and the complexity of the comparison of the individual terms is linear in the number
of terms. Thus, formulated as algebraic models, their difference becomes clear, a fact that otherwise might go
unnoticed and lead to discrepancies in model dynamics. 

Questions about global dynamical behavior can now be formulated in terms of
solving systems of polynomial equations, which come naturally from the
algebraic model. 
Due to the large number of variables involved in the above example, the resulting
polynomial systems can not be solved easily by direct computation on a standard
desktop computer with today's
methods and software. However, we believe that with the
advancement of algorithms and computational techniques, it will very soon be
possible to analyze such models with symbolic computation techniques.

Currently, there exist no feasible mathematical methods to analyze the
dynamics of general large stochastic models without iterating over the entire
state space, i.e., simulating the system, in whatever format. However, it is possible in many
cases to approximate a stochastic model with a deterministic system that
captures the main dynamical features. Next we show two examples where
mathematical theory can be used to analyze the dynamics of agent based
systems. 

\subsection{Simple Infection Model}
\begin{figure}[h!tp]
  \begin{center}
    \subfigure[Grid with a cell (black) and its 4 neighbors]
      {\label{fig:gridVirus}\includegraphics[width=.25\textwidth]{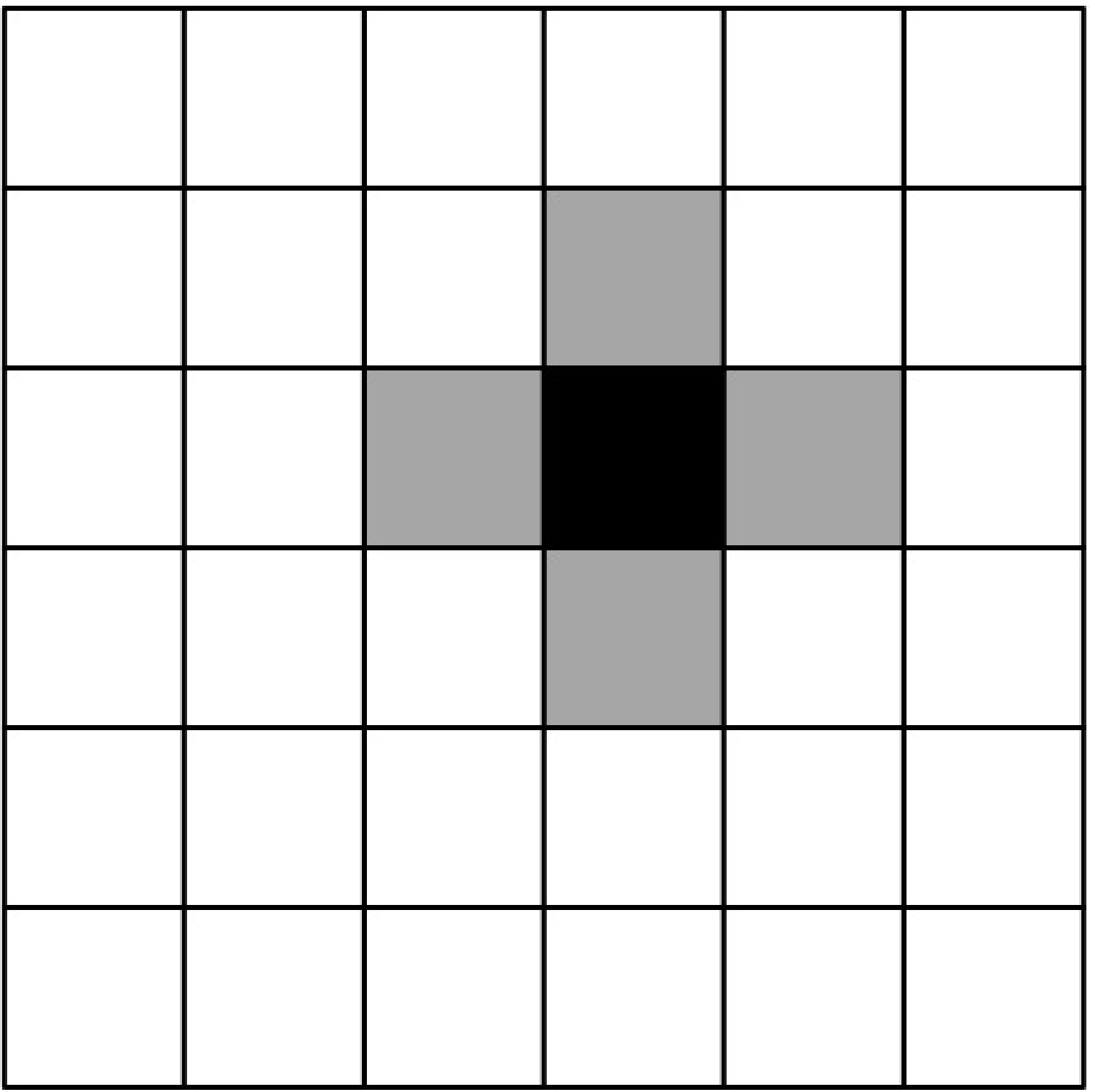}}
      \hspace{1in}
    \subfigure[Random infection of cells and the grid after one iteration]
      {\label{fig:grid12}\includegraphics[width=.25\textwidth]{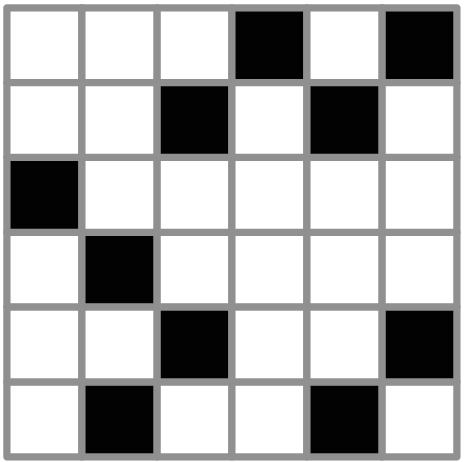}
        \includegraphics[width=.25\textwidth]{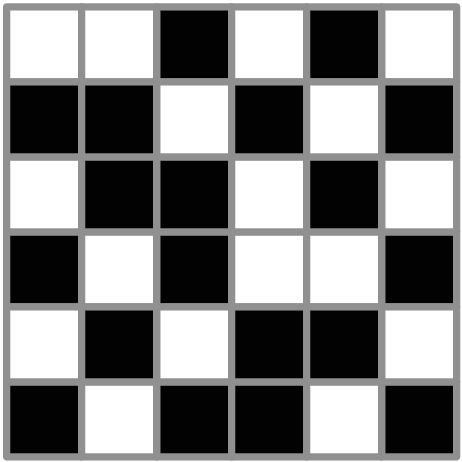}
      }
    \caption{Infection model}
  \end{center}
\end{figure}

Consider the following model for the spread of an infection: agents are cells on a square grid, every
cell can be healthy or infected. Healthy cells are depicted in white, infected
cells in black.
Figure \ref{fig:gridVirus} shows the layout of the grid with a cell and its four neighbors. Here, we
are considering the so-called von Neumann neighborhood of a cell. 

The systems evolves according to the following rules: A cell acquires a
healthy state, if its four neighbors are healthy, otherwise infected. Figure
\ref{fig:grid12} shows a randomly infected grid and its state after one
iteration.

\begin{figure}[h!tp]
  \begin{center}
  \includegraphics[width=.4\textwidth]{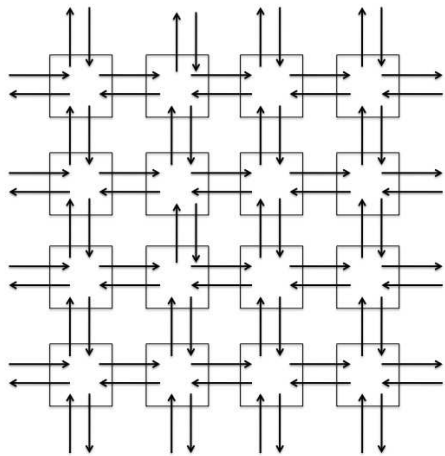}
  \end{center}
  \caption{Dependency graph for infection model}
  \label{fig:virusdep}
\end{figure}

This agent based model can easily be translated into an algebraic model.
Variables represent the cells, healthy cells have state ON (1), infected cells are
OFF (0). Since there are two states for each variable, we choose $\mathbb F_2$
as base field. The update rule is homogeneous for all cells, and for a cell $x$
with neighbors $y_1, y_2, y_3$, and $y_4$ it is given by $$f_x(y_1, y_2, y_3, y_4) =
y_1 y_2 y_3 y_4.$$ One easily sees that $f_x = 0$, i.e., infected, unless all
four neighbors are healthy ($1$). The resulting polynomial dynamical system
 is a so-called conjunctive Boolean network, as described
in \cite{ALAN}. The dependency graph of this network is shown in Figure
\ref{fig:virusdep} and clearly consists of one strongly connected component, that is,
any node can be reached from any other node by a directed path. We briefly
describe one of the results in \cite{ALAN} that applies here. 

\begin{figure}[h!tb]
  \begin{center}
  \includegraphics[width=.4\textwidth]{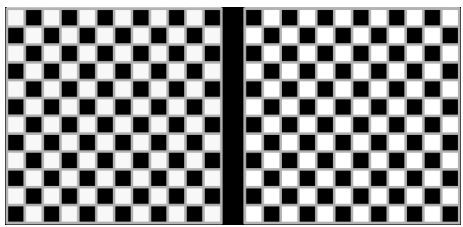}
  \end{center}
  \caption{Two states of periodicity 2}
  \label{fig:limitcycle}
\end{figure}

The {\it loop number} of a directed, strongly connected graph is defined as follows: 
choose a vertex (the loop number is independent of the vertex chosen) and consider all
directed loops at this vertex. The loop number is the greatest common divisor
of the lengths (number of edges) of all such loops \cite{ALAN}. It is easily seen that the infection
model has loop number $2$. Theorem 3.8 in \cite{ALAN} states that the length
of any limit cycle of the system has to divide the loop number, so that there are only
steady states and limit cycles of period 2. Furthermore, the theorem gives an example
formula for the number of limit cycles of a given length according to which the
system has exactly two steady states and one limit cycle of length 2, and
no other limit cycles. The two states of periodicity 2 are shown in
figure \ref{fig:limitcycle}. By the nature of the update rule the disease is
fast spreading, a single infected neighbor is sufficient for a cell to be
infected. It is interesting and counterintuitive that there is a limit
cycle of length 2 with only half the cells infected. The basin of attraction of this cycle
for an $n \times n$ grid is of size $2^{ \frac {n^2} 2 +1 }-2$, any
combination of at least half the cells being healthy and arranged as in Figure
\ref{fig:limitcycle}. 
There are two obvious steady states, given by all cells healthy or all cells infected. Although it
might be easy to find the two steady states and the limit cycle by studying the
agent based model and not its algebraic representation, one would still need
to prove that these are the only fixed points and that there is exactly one
limit cycle of length 2, and no limit cycles of greater length. 
Finding the fixed points for such a system by
simulation is computationally not feasible: on a $500$ by $500$ grid, there
are $2^{250000}$ different states in the state space that one would have to
test. 

The beauty of this algebraic model is, that the theorem in \cite{ALAN} applies
to an arbitrarily large finite grid. An elegant way of dealing with boundary cells
is to project the grid onto a torus by connecting the cells on the right edge to their
counterparts on the left edge, and similarly for the top and bottom cells.

\subsection{Conway's Game of Life}
Our third example is Conway's Game of Life \cite{ConwaysLife}, a 2-dimensional cellular automaton (CA),
using the 8 neighbors of a cell, 
that is, the Moore neighborhood of a cell,
to compute the next state.
While most ABMs are not in the form of a cellular automaton, Conway's Game of Life has some
of the same characteristics as many ABMs. It is also interesting for our purpose in that it poses 
an interesting computational challenge for our framework. The rules of this CA have to cover many cases
so that the polynomials expressing the rules are very dense, containing almost all possible terms.
This affects the computational complexity of the algorithms significantly.

Cells are in one of two states, either LIVE ($1$) or DEAD ($0$), with the following behavioral rules: 
\begin{enumerate}
  \item Any live cell with fewer than two live neighbors dies, as if caused by underpopulation.
  \item Any live cell with more than three live neighbors dies, as if by overcrowding.
  \item Any live cell with two or three live neighbors lives on to the next generation.
  \item Any dead cell with exactly three live neighbors becomes a live cell.
\end{enumerate}
Although the dynamics of the system are determined by very
simple rules, the emerging patterns are fascinating and have been studied
extensively. Questions that are natural to ask are what steady states or
oscillators can occur. We will show how to answer these
questions by using an algebraic model of the Game of Life. 

Naturally, the variables or agents in this system are the cells. 
There are only 2 possible states for an agent, DEAD or ALIVE. Therefore, we can
describe its behavior with polynomials over $\mathbb F _2$. Every agent $x$
has 8 neighbors, $x_1, \ldots, x_8$. The function $f_x$ that describes the
transition of agent $x$, is 
$$f_x(x, x_1, \ldots, x_8) = \left\{
  \begin{array}{ll}
    0 & : \sum x_i < 2\\
    0 & : \sum x_i = 2 \text{ and } x = 0\\
    1 & : \sum x_i = 2 \text{ and } x = 1\\
    1 & : \sum x_i = 3\\
    0 & : \sum x_i > 3\\
  \end{array}
\right.
$$
For function $f_x$ in polynomial form, see Appendix
\ref{app:polynomial}. The algebraic model for the Game of Life is a system 
$$f =(f_1,\dots,f_{n \times n}): \mathbb F_2^{n \times n} \longrightarrow \mathbb F_2^{n \times n}$$
$$x_i \mapsto f_i(x_1, \ldots, x_{n \times n}), $$
where $n$ is the dimension of the square grid. 
To calculate all the fixed points, i.e., still lives of
this system, we solve the system of polynomial equations: 
$$f_i(x) - x_i = 0, \qquad i = 1 \ldots n \times n.$$
One can use a computer algebra system like \emph{Macaulay2} \cite{M2} to compute a
Gr\"obner basis of the ideal that is generated by the equations for the fixed
points of the system, from which one obtains the fixed points. 

For example, on a $4 \times 4$ grid with periodic boundary conditions, the
fixed points of the model are the solutions to the system 
\begin{eqnarray*}
  f_1(x) &= x_1 \\
    \vdots\\
  f_{16}(x) &= x_{16}. \\
\end{eqnarray*}
First we compute a Groebner basis in lexicographic order for 
$I = \langle f_1(x) - x_1, \ldots, f_{16}(x) - x_{16}\rangle $
over the quotient ring 
$\mathbb F_2[x_1, \ldots, x_{16}]/J, $
where 
$J = \langle x^2_1 - x_1, \ldots, x^2_{16} - x_{16}\rangle$. From a
factorization of the Gr\"obner basis, one can easily read off the solutions. There are 53 fixed points. 

One should remark, that all fixed points can easily be found by updating every
possible initialization and checking whether it is a fixed point. 
For a $4 \times 4$ grid, there are only $2^{16} = 65536$ states, so finding
all fixed points is no computational challenge that would require computer
algebra. As the grid increases, the complexity of this brute force approach
increases exponentially, whereas the number of variables and equations
increases linearly in the algebraic model. 
For example, for a $10 \times 10$
grid one has to check $2^{100}$ states. With the algebraic model
though, one has to compute a Gr\"obner basis in a ring with 100 indeterminates
for an ideal with 100 generators. Depending on the polynomials describing the
update rules, this is a fast computation. 
Admittedly, for the functions that
describe the Game of Life, we were not able to compute the
solutions for a system with more than $16$ variables, neither with \emph{Macaulay2}
\cite{M2} nor \emph{Singular} \cite{GPS}. However, 
with the rapid improvement of hardware and computer algebra software, solving the
polynomial system to analyze the dynamics will soon become feasible for
larger grid sizes and also
faster than simulating every state which requires exhaustive enumeration of the state space. 

\section{Discussion}
At this time there is no broadly agreed-upon mathematical framework 
which can serve as a standard for the specification and analysis of agent-based models and
which preserves the key feature of this model type that global dynamics emerges from
local interactions. In this paper we propose such a framework, which preserves all
features of agent-based models and provides access to mathematical analysis tools. 
It is conceived as an extra step in the framework of the ODD protocol, which represents
a first step toward a standardized protocol for the specification of agent-based models.
The extra step can be automated, so that users do not need familiarity with the underlying
mathematical concepts.
The mathematical framework is that of polynomial dynamical systems over a finite field,
which provides access to theoretical and computational tools from computer algebra 
and discrete mathematics. 

We have presented examples of how this extra step of model specification works in practice,
and we have presented examples of how the mathematical specification provides added
value by allowing access to theoretical and computational tools for model analysis. We emphasize that
algebraic model specification is an addition to the ODD protocol, not a replacement. The
model must be explained in ODD to be understood by others. An algebraic system
is an additional resource that can be used to distribute and re-use the model. It eliminates any ambiguity
created by a verbal description, and it is a compact format that can run on
any system independent of software implementation, so parameters and rules
are easily modified for further simulations. The algebraic representation
allows easy comparison between two models. The rigorous mathematical language is another
advantage of the framework, rich algorithmic theory from computer
algebra and the theoretical foundation of algebraic geometry are available to
analyze algebraic models. We are making available a basic tool to automatically generate a
polynomial from data in ODD \cite{Parser} to ease the process of creating an
algebraic model. Because of its many advantages, we hope that modelers will
extend their model description to
the algebraic description and store their models in a central location so that
models can easily be found and re-used. 
The polynomial systems framework unifies the representation of three important discrete model types:
agent-based models, logical models (including Boolean network models), and Petri net models, allowing
direct comparison of different models. 

The translation algorithm presented in this paper only applies to deterministic agent-based models. And
the algorithms in \cite{Alan:Bioinf2010} also deal only with deterministic models. However, the majority of published ABMs in biology are stochastic. Fortunately, there are stochastic versions of several of these 
model types available on the mathematical
side, that can be used for the purpose of modeling stochastic ABMs. The most suitable model type is that 
of probabilistic Boolean networks \cite{pbn-shmulevich}, a multi-state generalization of 
Boolean networks that is stochastic in two ways: each node of the network has attached a probability space of update functions rather 
than a single function and, secondly, the update order can be stochastic. It is easy to represent these models
in the polynomial dynamical system framework, which we have done as part of our software package
ADAM (Analysis of Dynamic Algebraic Models), available at
\href{http://adam.vbi.vt.edu}{http://adam.vbi.vt.edu} as a web service.

Finally, we comment on the computational complexity of the analysis of agent-based models by the methods
proposed in this paper. While a significant number of published ABMs are well within the computational reach of our methods, there are many ABMs that completely overwhelm them. This is of course similar to the situation
for continuous models and parameter estimation methods, bifurcation analysis, etc. There too, models
are becoming too large to be analyzed by anything other than more or less through simulation. For some ABMs
even simulation represents a challenge because of their size. In these cases 
new methods for model reduction are the only
viable approach to a mathematical analysis, no matter what methods are available. 
For instance, in the case of a multi-scale ABM one possible approach might be to construct
phenomenological models for the higher scales that accurately model the aggregate dynamics of
the lower scales without explicitly representing these. This is the subject of ongoing work by the authors. 

\newpage
\appendix
\section{Translation from ODD to polynomial model}\label{app:construction}

Here we describe the details of constructing a polynomial model from ABM information stored in the
ODD protocol format. 

\subsection{Update schedule} 
The scheduling information provided allows the assembly of a complete update order for all the
state variables that need to be updated, possibly involving a mixture of sequential, parallel, deterministic
and random updates of subgroups of the state variables. The scheduling information can be assembled
to a probability space $P$, which has as elements the set of words in the letters 
$u_1,\ldots , u_n; v_1,\ldots , v_n$. Each word $u_iu_ju_k\cdots v_av_b\cdots u_c\cdots $ translates
into an update order $x_ix_jx_k\cdots (x_ax_b\cdots )x_c\cdots $, which is to be interpreted as
updating first $x_i, x_j, x_k, \ldots $ sequentially in this order, then updating $x_a, x_b, x_c, \ldots$
in parallel, then update $x_c,\ldots $ sequentially in this order, etc. The probability distribution on the
space $P$ can be computed from the scheduling information which indicates the
variables to be updated randomly and with which probability distribution. The resulting dynamical system
will be denoted as
$$
f = (P_1,\ldots , P_n;P): X\longrightarrow X.
$$

At this point we have described the dynamical system as a set function, which is rather limiting. 
For instance, if we want to compare whether two such models are identical, we need to evaluate both
at all possible inputs and compare the outputs. It would be useful if we could describe the function $f$
via equations. Now, sometimes, the model specification will already be given to us in the form of 
equations or mathematical functions. We now describe a procedure that allows us to represent $f$ by
mathematical functions of a unified type, in all circumstances. This procedure is analogous to the
introduction of a Cartesian coordinate system to transition from synthetic geometry to analytic geometry.
The fundamental observation is the following.

\subsubsection{Observation} \label{sec:observation} 
Let $\mathbb F$ be a finite field, such as $\mathbb Z/p$ and let 
$f: \mathbb F^n\longrightarrow \mathbb F$ be any function. 
Then there exists a unique polynomial $g\in \mathbb F[x_1, \ldots , x_n]$, such that each variable in $g$ appears
to a power less than $|\mathbb F|$, and $f(a_1,\ldots , a_n) = g(a_1,\ldots , a_n)$ for all 
$(a_1, \ldots , a_n)\in \mathbb F^n$. 

Thus, in the case that all state variables take on states in the same state set, which furthermore can
be given the structure of a finite field, then the dynamical system $f: X\longrightarrow X$ above can
be described via polynomials. This is in fact always possible, and we briefly outline the process. It is 
similar to a construction described in detail in \cite{Alan:Bioinf2010}. There, we show how
to translate a so-called logical model into a polynomial dynamical system. Logical models are
specified in a way that is very similar to the rules for state variables in an ABM. First consider one state
variable and suppose that one of its states is described by an $r$-tuple from a set 
$X_1\times\cdots\times X_r$. In each component $X_i$ we choose an element and duplicate it
enough times so that the number of elements in $X_i$ becomes equal to the smallest prime number
that is greater than or equal to the orders of all the $X_j$. If $X_i$ contains ordered numerical values,
for instance, say $X_i = \{1, 2, 3, 4\}$, then we can add a state $4'$ to obtain a set with 5 elements.
After carrying this construction out for all $X_i$ we have obtained a set $X_1\times\cdots\times X_r$
in which each component has the same number of elements, equal to the power of a prime number $p$. It is now
straightforward to see that one can endow this set with a field structure so that it becomes
isomorphic to the Galois field $\mathbb F_{p^r}$. A similar construction can now be carried out to assure that the order $p^r$ is the same for the Galois fields for all state variables. The end result is that all state
variables take values in the same finite field $\mathbb F$.
The last step is to extend the update functions $f_i$ for each state variable to the larger state set. This is done
by assigning the same value to a new state as the state that was duplicated to obtain it.

\section{Behavioral Rule in Polynomial Form}\label{app:polynomial}
For a $4$ by $4$ grid, we obtain for agent
$x_1$ with neighbors $x_2\ldots, x_9$ the following polynomial: \\
$f =
x_1*x_2*x_3*x_4*x_5*x_6*x_7*x_8+x_1*x_2*x_3*x_4*x_5*x_6*x_7*x_9+x_1*x_2*x_3*x_4*x_5*x_6*x_8*x_9+x_1*x_2*x_3*x_4*x_5*x_7*x_8*x_9+x_1*x_2*x_3*x_4*x_6*x_7*x_8*x_9+x_1*x_2*x_3*x_5*x_6*x_7*x_8*x_9+x_1*x_2*x_4*x_5*x_6*x_7*x_8*x_9+x_1*x_3*x_4*x_5*x_6*x_7*x_8*x_9+x_1*x_2*x_3*x_4*x_5*x_6*x_7+x_1*x_2*x_3*x_4*x_5*x_6*x_8+x_1*x_2*x_3*x_4*x_5*x_7*x_8+x_1*x_2*x_3*x_4*x_6*x_7*x_8+x_1*x_2*x_3*x_5*x_6*x_7*x_8+x_1*x_2*x_4*x_5*x_6*x_7*x_8+x_1*x_3*x_4*x_5*x_6*x_7*x_8+x_2*x_3*x_4*x_5*x_6*x_7*x_8+x_1*x_2*x_3*x_4*x_5*x_6*x_9+x_1*x_2*x_3*x_4*x_5*x_7*x_9+x_1*x_2*x_3*x_4*x_6*x_7*x_9+x_1*x_2*x_3*x_5*x_6*x_7*x_9+x_1*x_2*x_4*x_5*x_6*x_7*x_9+x_1*x_3*x_4*x_5*x_6*x_7*x_9+x_2*x_3*x_4*x_5*x_6*x_7*x_9+x_1*x_2*x_3*x_4*x_5*x_8*x_9+x_1*x_2*x_3*x_4*x_6*x_8*x_9+x_1*x_2*x_3*x_5*x_6*x_8*x_9+x_1*x_2*x_4*x_5*x_6*x_8*x_9+x_1*x_3*x_4*x_5*x_6*x_8*x_9+x_2*x_3*x_4*x_5*x_6*x_8*x_9+x_1*x_2*x_3*x_4*x_7*x_8*x_9+x_1*x_2*x_3*x_5*x_7*x_8*x_9+x_1*x_2*x_4*x_5*x_7*x_8*x_9+x_1*x_3*x_4*x_5*x_7*x_8*x_9+x_2*x_3*x_4*x_5*x_7*x_8*x_9+x_1*x_2*x_3*x_6*x_7*x_8*x_9+x_1*x_2*x_4*x_6*x_7*x_8*x_9+x_1*x_3*x_4*x_6*x_7*x_8*x_9+x_2*x_3*x_4*x_6*x_7*x_8*x_9+x_1*x_2*x_5*x_6*x_7*x_8*x_9+x_1*x_3*x_5*x_6*x_7*x_8*x_9+x_2*x_3*x_5*x_6*x_7*x_8*x_9+x_1*x_4*x_5*x_6*x_7*x_8*x_9+x_2*x_4*x_5*x_6*x_7*x_8*x_9+x_3*x_4*x_5*x_6*x_7*x_8*x_9+x_1*x_2*x_3*x_4+x_1*x_2*x_3*x_5+x_1*x_2*x_4*x_5+x_1*x_3*x_4*x_5+x_1*x_2*x_3*x_6+x_1*x_2*x_4*x_6+x_1*x_3*x_4*x_6+x_1*x_2*x_5*x_6+x_1*x_3*x_5*x_6+x_1*x_4*x_5*x_6+x_1*x_2*x_3*x_7+x_1*x_2*x_4*x_7+x_1*x_3*x_4*x_7+x_1*x_2*x_5*x_7+x_1*x_3*x_5*x_7+x_1*x_4*x_5*x_7+x_1*x_2*x_6*x_7+x_1*x_3*x_6*x_7+x_1*x_4*x_6*x_7+x_1*x_5*x_6*x_7+x_1*x_2*x_3*x_8+x_1*x_2*x_4*x_8+x_1*x_3*x_4*x_8+x_1*x_2*x_5*x_8+x_1*x_3*x_5*x_8+x_1*x_4*x_5*x_8+x_1*x_2*x_6*x_8+x_1*x_3*x_6*x_8+x_1*x_4*x_6*x_8+x_1*x_5*x_6*x_8+x_1*x_2*x_7*x_8+x_1*x_3*x_7*x_8+x_1*x_4*x_7*x_8+x_1*x_5*x_7*x_8+x_1*x_6*x_7*x_8+x_1*x_2*x_3*x_9+x_1*x_2*x_4*x_9+x_1*x_3*x_4*x_9+x_1*x_2*x_5*x_9+x_1*x_3*x_5*x_9+x_1*x_4*x_5*x_9+x_1*x_2*x_6*x_9+x_1*x_3*x_6*x_9+x_1*x_4*x_6*x_9+x_1*x_5*x_6*x_9+x_1*x_2*x_7*x_9+x_1*x_3*x_7*x_9+x_1*x_4*x_7*x_9+x_1*x_5*x_7*x_9+x_1*x_6*x_7*x_9+x_1*x_2*x_8*x_9+x_1*x_3*x_8*x_9+x_1*x_4*x_8*x_9+x_1*x_5*x_8*x_9+x_1*x_6*x_8*x_9+x_1*x_7*x_8*x_9+x_1*x_2*x_3+x_1*x_2*x_4+x_1*x_3*x_4+x_2*x_3*x_4+x_1*x_2*x_5+x_1*x_3*x_5+x_2*x_3*x_5+x_1*x_4*x_5+x_2*x_4*x_5+x_3*x_4*x_5+x_1*x_2*x_6+x_1*x_3*x_6+x_2*x_3*x_6+x_1*x_4*x_6+x_2*x_4*x_6+x_3*x_4*x_6+x_1*x_5*x_6+x_2*x_5*x_6+x_3*x_5*x_6+x_4*x_5*x_6+x_1*x_2*x_7+x_1*x_3*x_7+x_2*x_3*x_7+x_1*x_4*x_7+x_2*x_4*x_7+x_3*x_4*x_7+x_1*x_5*x_7+x_2*x_5*x_7+x_3*x_5*x_7+x_4*x_5*x_7+x_1*x_6*x_7+x_2*x_6*x_7+x_3*x_6*x_7+x_4*x_6*x_7+x_5*x_6*x_7+x_1*x_2*x_8+x_1*x_3*x_8+x_2*x_3*x_8+x_1*x_4*x_8+x_2*x_4*x_8+x_3*x_4*x_8+x_1*x_5*x_8+x_2*x_5*x_8+x_3*x_5*x_8+x_4*x_5*x_8+x_1*x_6*x_8+x_2*x_6*x_8+x_3*x_6*x_8+x_4*x_6*x_8+x_5*x_6*x_8+x_1*x_7*x_8+x_2*x_7*x_8+x_3*x_7*x_8+x_4*x_7*x_8+x_5*x_7*x_8+x_6*x_7*x_8+x_1*x_2*x_9+x_1*x_3*x_9+x_2*x_3*x_9+x_1*x_4*x_9+x_2*x_4*x_9+x_3*x_4*x_9+x_1*x_5*x_9+x_2*x_5*x_9+x_3*x_5*x_9+x_4*x_5*x_9+x_1*x_6*x_9+x_2*x_6*x_9+x_3*x_6*x_9+x_4*x_6*x_9+x_5*x_6*x_9+x_1*x_7*x_9+x_2*x_7*x_9+x_3*x_7*x_9+x_4*x_7*x_9+x_5*x_7*x_9+x_6*x_7*x_9+x_1*x_8*x_9+x_2*x_8*x_9+x_3*x_8*x_9+x_4*x_8*x_9+x_5*x_8*x_9+x_6*x_8*x_9+x_7*x_8*x_9$.
\newline
\pagebreak
\bibliographystyle{plain}
\bibliography{control}
\end{document}